\documentclass[times]{anzsauth}
\usepackage{moreverb}
\usepackage[colorlinks,bookmarksopen,bookmarksnumbered,citecolor=red,urlcolor=red]{hyperref}
\usepackage{hyperref}
\newcommand\BibTeX{{\rmfamily B\kern-.05em \textsc{i\kern-.025em b}\kern-.08em
T\kern-.1667em\lower.7ex\hbox{E}\kern-.125emX}}

\def\volumeyear{2015}

\begin{document}

 \runningheads{Bayesian clustering of action potential spikes}{Z.~van~Havre,~N. White,~J.~Rousseau,~K.~Mengersen}

\title{Clustering action potential spikes:\\ Insights on the use of overfitted finite mixture models and Dirichlet process mixture models.}

\author{Z.~van~Havre\corrauth\footnotemark[2]\footnotemark[3]\footnotemark[4],~N.~White\footnotemark[2],~J.~Rousseau\footnotemark[3],~and~K.~Mengersen\footnotemark[2]}
\affiliation{ARC Centre of Excellence for Mathematical and Statistical Frontiers (ACEMS), Queensland University of Technology \& \\ Centre De Recherche en Math\'{e}matiques de la D\'{e}cision (CEREMADE),  Universit\'{e} Paris-Dauphine}

\footnotetext[1]{zoevanhavre@gmail.com}
\footnotetext[2]{ARC Centre of Excellence for Mathematical and Statistical Frontiers (ACEMS), Queensland University of Technology, Brisbane, Australia }
\footnotetext[3]{Centre De Recherche en Math\'{e}matiques de la D\'{e}cision (CEREMADE),  Universit\'{e} Paris-Dauphine, Paris, France}
\footnotetext[4]{CSIRO Health and Biosecurity/Australian E-Health Research Centre, Brisbane QLD, Australia}

\begin{abstract}
The modelling of action potentials from extracellular recordings, or spike sorting, is a rich area of neuroscience research in which latent variable models are often used. Two such models, Overfitted Finite Mixture models (OFMs) and  Dirichlet Process Mixture models (DPMs) are considered to provide insights for  unsupervised clustering of complex, multivariate medical data when the number of clusters is unknown. OFM and DPM are structured in a similar hierarchical fashion but they are based on different philosophies with different underlying assumptions. This study investigates how these differences impact on a real study of spike sorting, for the estimation of multivariate Gaussian location-scale mixture models  in the presence of  common difficulties arising from complex medical data. The results provide insights allowing the future analyst to choose an approach suited to the situation and goal of the research problem at hand.
\end{abstract}

 \keywords{Spike sorting; Bayesian inference; Mixtures; Multivariate Gaussian; Unsupervised clustering; Latent class model; Noise; Outliers; Parkinson's Disease}
 
\maketitle

\section{Introduction}\label{sec:Intro}
Extracellular recordings are an indispensable tool in neuroscience, enabling the real time monitoring of neural activity.  Central to these recordings is the measurement of action potentials (APs) or ``spikes'', which provide an indication of neuron populations present in the region of interest.  Characterisation of these can further our understanding of various neural mechanisms, including responses to various stimuli.

Spike sorting refers to the collection of techniques suited to this purpose, encompassing the stages of AP detection, processing and classification.  Comprehensive reviews of each of these stages are provided by \citet{Lewicki1998} and \citet{Sahani1999}.  This paper focuses on the final step, corresponding to inference around the assignment of individual spikes to source neurons.  Common to approaches developed for this purpose is the grouping of APs based on templates or a reduced set of features, for example, firing statistics \citep{Delescluse2006,Pouzat2004}, wavelet transforms \citep{letelier2000} or principal components \citep{Lewicki1998} .

Mixture models are a popular tool for AP classification by virtue of their unsupervised approach to clustering, and are used to group individual APs into clusters containing spikes of similar shapes representing different sources of neurological activity.  These models are underpinned by the assumption that each AP has been generated by one of a number of distinct clusters, where the composition of each cluster is \textit{a priori} unknown.  To date, methodologies proposed include finite mixtures of Gaussian \citep{Sahani1999,Wood2004} and t-distributions \citep{Shoham2003}, mixtures of factor analysers \citep{gorur2004spike}, Reversible Jump Markov chain Monte Carlo (RJMCMC) \citep{nguyen2003}, and time-dependent mixtures to account for non-stationarity \citep{Bar-Hillel2006,calabrese2011kalman}.  Non-parametric approaches to mixture estimation, based on the Dirichlet Process, have also been posited \citep{Wood2008,Gasthaus2009}.
Whilst these models share the common goal of clustering, there exist key differences in the assumptions underpinning their development.  This has the potential to impact on subsequent model-based inferences, particularly when interest lies in the determination of the optimal clustering of the observed data.  The implications of applying different mixture-based methods, in the context of the featured problem, remains relatively unexplored.

This paper seeks to provide insight into this issue by comparing two mixture-based approaches to spike classification, both of which are formulated within the Bayesian framework.  The first model considered is a finite mixture of multivariate Gaussian distributions, applying methodology recently proposed by \citet{VanHavre2015}.  Outlined further in Section \ref{sec:OFmodel}, this method approaches mixture model estimation by initially overfitting the number of clusters expected to be present.  Specified conditions on the prior distribution for the mixture weights encourage the emptying out of excess components in the posterior distribution \citep{Mengersen2011}.  The performance of this methodology is compared to a Dirichlet Process mixture model, a non-parametric alternative to mixture modelling.

The chosen models are compared with respect to three objectives.  Firstly, differences in posterior inference concerning the number of occupied clusters are compared.  The second objective seeks to reconcile differences in the inferred classification of the APs into clusters, based on posterior pairwise probabilities of individual APs being assigned to the same cluster.  Finally, the third objective is to compare the composition of the mixture components, under an inferred, optimal cluster configuration.  To address these objectives, the models are applied to spikes collected from the subthalamic nucleus during Deep Brain Stimulation, a surgical intervention for the alleviation of symptoms in patients with advanced Parkinson`s disease (PD).

The remainder of this paper is organised as follows. In Section \ref{sec:data}, the data used for analysis are described.  Key details of the chosen mixture models are outlined in Section \ref{sec:methods}.  The results of the aforementioned application are reported in Section \ref{sec:results}, and are organised in line with the stated objectives.  A discussion of key results and directions for future work are summarised in Section \ref{sec:discussion}.

\section{Data and Methodology}

\subsection{Data}\label{sec:data}
\begin{figure}[htbp!]
\centering
\includegraphics[width=\textwidth]{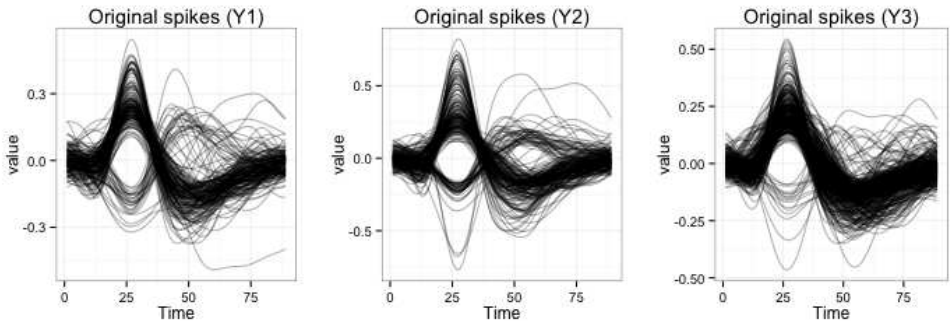}
\caption[Sampled waveforms detected in each dataset ($Y_1$,$Y_2$,$Y_3$).]{Sampled waveforms detected in each dataset ($Y_1$,$Y_2$,$Y_3$).}
\label{Figure_1}
\end{figure}

The data analysed in this paper consisted of extracellular recordings of the subthalamic nucleus collected during Deep Brain Stimulation, a treatment for advanced Parkinson's Disease.  These data are comprised of spikes extracted from three independent recordings of Deep Brain Stimulation, using standard techniques, and was originally analysed by \citet{White2011a} (see \citet{White2011a} for further details).  The resulting waveforms from each recording, labelled $Y_1$, $Y_2$ and $Y_3$ are displayed in Figure \ref{Figure_1}, with sample sizes $n=192$, $211$ and $348$, respectively.

\begin{figure}[htbp!]
\centering
 \includegraphics[width=\textwidth]{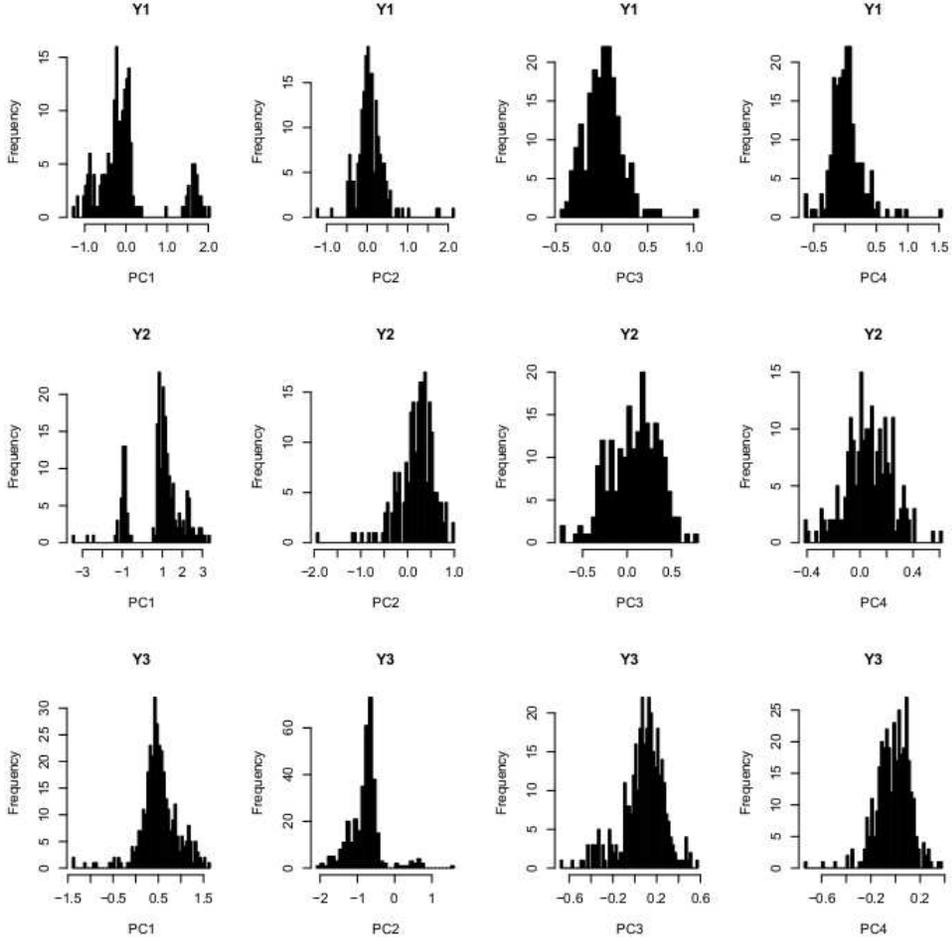}
\caption[Histogram of the first four principal components of each set of spike data]{Histogram of the first four principal components of each set of spike data.}
\label{Figure_2}
\end{figure}

Dimension reduction was performed on the sampled waveforms using a robust version of Principal Components Analysis (PCA) \citep{hubert2005robpca}, to lessen the influence of outliers on the calculation of principal components (PCs).  For all analyses presented in Section \ref{sec:results}, the first four principal components (PCs) were used as inputs into each model for all three datasets, in each case explaining greater than 80\% of variation among waveforms (83\%, 91\%, and 85\% for the first, second and third dataset respectively). Robust scree plots were used to assist this choice, and the four PC's are illustrated in Figure \ref{Figure_2}.

\subsection{Methodology}\label{sec:methods}
In this Section, the key details of the two proposed mixture-based approaches are outlined.  Common to both approaches is the problem of inferring the partition of $n$ (possibly multivariate) observations into $K$ clusters, where $K$ is unknown.  For the sample $\mathbf{y}=\{\mathbf{y}_1, \dots, \mathbf{y}_n\}$, let $\mathbf{y}_i=\{y_{i1}, \dots, y_{ir}\}$ consist of $r$ measurements associated with the $i$'th observation.  Cluster membership for each $\mathbf{y}_i$ is inferred via the discrete latent variable $z_i$, with $z_i=k$ denoting the assignment of $\mathbf{y}_i$ to cluster $k$.

In light of data described in Section \ref{sec:data}, assume that, conditional on assignment to a cluster $k$, $\mathbf{y}_i$ follows a Multivariate Gaussian distribution with mean $\boldsymbol\mu_k=\left[\mu_{1k},\ldots,\mu_{rk}\right]$ and variance-covariance matrix $\Sigma_k$, $1\leq k \leq K$.  Conditional on assignment to cluster $k$, the likelihood for $\mathbf{y}_i$ is,
\begin{equation}
p\left(\mathbf{y}_i|z_i=k,\boldsymbol\theta_k\right)=N_{r}\left(\boldsymbol\mu_k,\boldsymbol\Sigma_k\right),
\end{equation}
defined by the unknown parameter set $\boldsymbol\theta_k=\left(\boldsymbol\mu_k,\boldsymbol\Sigma_k\right)$.

In this paper, each model was estimated using Markov chain Monte Carlo (MCMC), with details provided in the relevant subsections.  For each $k$, a joint prior distribution of form $p\left(\boldsymbol\mu_k,\boldsymbol\Sigma_k\right)=p\left(\boldsymbol\mu_k|\boldsymbol\Sigma_k\right)p\left(\boldsymbol\Sigma_k\right)$ was adopted, with
\begin{align}\label{eq:mvnpriors}
p\left(\boldsymbol\mu_k|\boldsymbol\Sigma_k\right)&=\mathcal{N}_{r}\left(\mathbf{b}_0,\frac{\boldsymbol\Sigma_k}{N_0}\right)\nonumber \\
p\left(\Sigma_k\right)&=\mathcal{IW}\left(c_0,\mathbf{C}_0\right),
\end{align}
where $E[\boldsymbol\Sigma_k]=\frac{\mathbf{C}_0}{\left(c_0-r-1\right)}$. The choice of hyperparameters $\left(\mathbf{b}_0,N_0,c_0,\mathbf{C}_0\right)$ for the featured study is discussed in Section \ref{sec:results}.

\subsubsection{Overfitted Finite Mixture model (OFM)}\label{sec:OFmodel}
The Overfitted Finite Mixture model (OFM) approach consisted of fitting a finite mixture model where the number of components specified, say $K^*$, is greater than the true number of clusters.  Using the aforementioned notation, the likelihood of $\mathbf{y}$ is given by
\begin{equation}
p ( \mathbf{y} | \pmb{\theta},\pmb{\pi} )   = \prod_{i=1}^{n} \sum\limits_{k=1}^{K^*}  \pi_{k}N_{r}\left(\boldsymbol\mu_k,\boldsymbol\Sigma_k\right) , \label{eq:fmmlikey}
\end{equation}
where $K^*>K$ and $\pi_k = Pr\left(z_i=k\right)$, corresponding to the \textit{a priori} probability of a randomly selected observation being assigned to cluster $k$.  Collectively, $\pmb\pi=\{\pi_{1},\ldots,\pi_{K^*}\}$ represent the mixture weights and are subject to the constraint $\sum_{k=1}^{K^*}\pi_k=1$.

Given the above definition of the mixture weights, the prior distribution for each $z_i$ is Multinomial,
\begin{equation}
z_i|\pmb\pi\sim \mathcal{MN}\left(1;\pi_1,\ldots,\pi_{K^*}\right)
\label{eq:fmmzprior}
\end{equation}
and is updated as part of the chosen MCMC scheme.  The inclusion of $z_i$ as a latent variable represents a form of data augmentation \citep{Tanner1987}, to facilitate feasible computation of Equation (\ref{eq:fmmlikey}).

The key feature of the OFM approach lies in the choice of prior distribution for the mixture weights.  In this case, the vector of weights follows a Dirichlet distribution,
\begin{equation}
(\pi_{1}, \dots ,\pi_{K^*}) \sim \mathcal{D}(\alpha_1,\dots, \alpha_{K^*}),
\end{equation}
characterised by the hyperparameter $\alpha_k$ for $k=1,\dots, {K^*}$. We consider here the case $\alpha_1=\dots=\alpha_{K^*}$ to obtain an exchangeable prior.

Building on theoretical results first published by \cite{Mengersen2011}, overfitting finite mixture models with posterior emptying has recently been rendered achievable using well-known MCMC techniques \citep{VanHavre2015}. The methodology (called Zmix), was applied by \citet{VanHavre2015} to the case of univariate Gaussian mixture models.
 The OFM depends on the choice of an appropriate value of $\alpha$ that results in the excess components $\{k: K<k\ \leq K^*\}$ being assigned negligible weight, so that in practice no observations are allocated to unnecessary groups. As a result, the partition of $\mathbf{y}$ implied by $\mathbf{z}$ provides insight into the true number of mixture components. 
The appropriate choice of $\alpha_k$ was considered at length in \citep{VanHavre2015},  who showed that when $K$ is unknown, a very small value of $\alpha_k$ close to zero must be used, to prevent extra groups from being populated in practice.

The resulting posterior parameter surface is difficult for a Gibbs sampler to traverse thanks to the complex mutlimodal nature of OFMs, which is exacerbated by this choice of hyperparameter. The MCMC can however be augmented with a Prior Parallel Tempering algorithm, which enables accurate estimation of the OFM posterior with plentiful mixing by drawing on the behaviour of the OFM under small changes in the prior hyperparameters \citep{VanHavre2015}. 

Briefly, writing $\pmb{\alpha}^{(t)}=\{\alpha_1, \dots, \alpha_K\}^{(t)}$, a ladder of $T$ values $\{ \pmb{\alpha}^{(1)},\dots, \pmb{\alpha}^{(T)} \}$  was created where  $\pmb{\alpha}^{(1)}$ corresponded to the smallest, chosen \textit{a priori} to promote emptying behaviour based on the results of \cite{Mengersen2011}. As $t$ increases, the values of $\pmb{\alpha}^{(t)}$ increase gradually, eventually reaching values which entirely prevent empty clusters from being estimated .

The MCMC algorithm is implemented in parallel in combination with Gibbs sampling steps for the remaining model parameters.  Further details of the Zmix algorithm can be found in \cite{VanHavre2015}.  Code used to implement the MCMC algorithm for the OFM model is provided in \nameref{Code_1_SuppInfo}.

\subsubsection{The Dirichlet Process Mixture model (DPM)}\label{sec:NPmodel}
Dirichlet Process mixture (DPM) modelling has become increasingly popular for unsupervised clustering, and is commonly viewed as a nonparameteric alternative to the finite mixture model.  DPMs are distinguished by their use of a Dirichlet Process (DP) as a prior over mixture model components.  Formally, for a given measurable space $\pmb{\Theta}$, the DP is a stochastic process  defined as a distribution over probability measures.  Its use as a prior distribution leads to the following  model for $\mathbf{y}_i$,
\begin{align}
\mathbf{y}_i|\pmb{\theta}_i&\sim \pmb{\theta}_i\nonumber\\
\pmb{\theta}_i|G&\sim G\nonumber\\
G&\sim\mathcal{DP}\left(m G_0\right).\label{eq:DP1}
\end{align}
where $G$ denotes a random probability measure. The DP in Equation (\ref{eq:DP1}) is defined by a base distribution, $G_0$, corresponding to the mean of the DP, and concentration parameter $m>0$.

The applicability of the DP within a mixture setting is attributed to its discreteness property, that states the existence of a non-zero probability of multiple $\theta_i$'s equalling the same value, in turn inducing clustering behaviour.  This discrete nature of $G$ is made clear with the stick-breaking construction \citep{sethuraman1994}, that replaces $G$ with an infinite weighted sum of point masses.  Using this construction, the DP is rewritten as,
\begin{align}
G&=\sum_{k=1}^{\infty}\pi_k\delta_{\boldsymbol\theta_{k}}\nonumber\\
\pi_{k}&=v_k\prod_{l<k}\left(1-v_l\right)\nonumber\\
v_k&\sim Beta\left(1,m\right)\nonumber\\
\theta_k|G_0&\sim G_0\label{eq:DPstick}{K^*}
\end{align}
where $G_0=p\left(\pmb{\theta}_k\right)$ and for each $k=1,2,\dots$, $\delta_{\pmb{\theta}_{k}}$ denotes a Dirac mass at $\theta_k$ weighted by $\pi_k$; $\sum_{k=1}^{\infty}\pi_k=1$. The term `stick-breaking' refers to the analogy that the weights $\pi_1,\pi_2,\ldots$ represent portions of a unit-length stick, with each $\pi_k$ being a randomly drawn proportion of length remaining, given preceding clusters.  Similar to the OFM, $\pi_1,\pi_2,\ldots$ represent cluster weights, albeit in this case, $K$ is countably infinite.  For this reason, the DPM is often referred to as an infinite mixture model \citep{teh2010}.

Compared to the OFM in Section \ref{sec:OFmodel}, the DPM introduces an additional parameter, $m$, related to the concentration of the DP around the base distribution.  In light of the stick-breaking construction in Equation (\ref{eq:DPstick}), $m$ is seen to influence the construction of the weights, $v_1,v_2,\dots$ that, in turn, directly influence each $\pi_k$.  In this paper, the true value of $m$ is assumed unknown and assigned a $\mathcal{G}\left(1,1\right)$ prior distribution.
This choice encourages a small number of clusters a priori, and is centered around 1, a commonly chosen value when $m$ is assumed to be fixed \citep{teh2010}.

Similar to the OFM, each $\mathbf{y}_i$ is associated with $\pmb{\theta}_k$ through the introduction of a latent variable $z_i$, which is updated as part of the MCMC scheme.  For the results presented in Section \ref{sec:results}, the slice sampler proposed by \citet{walker2007} was implemented. This algorithm proposes an efficient way of sampling from the stick-breaking construction via the introduction of auxiliary variables. The key benefit of this approach is that it avoids the need to approximate the DP by truncating the number of components computed \citep{ishwaran2001gibbs}.  Rather, the resulting slice sampler allows for adaptive truncation of the stick-breaking representation in a way that enables feasible computation of $\mathbf{z}$.  To improve the mixing behaviour of the sampler, two label switching moves were also implemented \citep{papaspiliopoulos2008}.  R code to implement the chosen MCMC scheme is provided in \nameref{Code_2_SuppInfo}.

Due to the countably infinite dimension of the DPM, the primary focus is not on the determination of the optimal value for $K$.  Rather, one seeks to infer similarities among elements of $\mathbf{y}$.  The most common method for achieving this inference is to compute posterior pairwise probabilities of equality for all pairs of labels $z_i$ and $z_{i'}$, i.e., $Pr\left(z_i=z_{i'}|\mathbf{y}\right)$ $\forall$ $i\neq i'$.  This approach is easily implemented within an MCMC framework and has the benefit of being label invariant.  Furthermore, in cases in which the optimal partition of $\mathbf{y}_i$ is sought, there exist a number of methods for the determination of the maximum a posteriori (MAP) estimate of $\mathbf{z}$, based on the $n\times n$ matrix of posterior pairwise probabilities \citep{medvedovic2002}.  For results presented in Section \ref{sec:results}, the Posterior Expected Rand (PEAR) index proposed by \citet{fritsch2009improved}, which is  based on the Rand index \citep{Rand1971}, was adopted for this purpose.

\section{Results}\label{sec:results}
The results of modelling the three multivariate datasets under both approaches (OFM and DPM) are explored according to the three objectives detailed in Section \ref{sec:Intro}.   For both the OFM and DPM, the results presented are based on 50,000 MCMC iterations, discarding the first 25,000. Convergence to the target distribution after burn-in was assessed via graphical examination of the MCMC samples. For the estimation of each OFM, $K^*=10$ components were included in the mixture model fit to each dataset.  Likewise, for the estimation of each DPM, the slice sampler was initialised with $K^*=10$ clusters. $K^*$ was chosen to be  larger than the maximum expected number of components contained in each dataset. In each case, the allocation variable $\mathbf{z}$ was initialised based on the results of a k-means algorithm on $\mathbf{y}$ (assuming $K^*$ clusters).

To specify the prior distributions defined in Equation \ref{eq:mvnpriors}, the following values were chosen for the hyperparameters: $\mathbf{b}_0=\overline{\mathbf{y}}$, $N_0=0.01$, $c_0=5$ and $\mathbf{C}_0=0.75\textrm{cov}\left(\mathbf{y}\right)$.  These values were chosen to reflect a plausible range of values for each parameter, whilst remaining relatively non-informative.  Similar choices for  multivariate Gaussian mixture models are discussed in \citet{Fruhwirth-Schnatter2006}.

As noted in Section \ref{sec:data}, the first four principal components computed explained a large proportion of the variation in the data and were extracted for this analysis, such that each resulting dataset has $r=4$ dimensions.  Throughout this Section, these datasets are referred to as $Y_1$, $Y_2$ and $Y_3$, respectively.

\subsection{Distribution of occupied clusters}\label{sec:Result_1}
Objective 1 is to compare the distribution of the number of occupied components in the results generated by the two approaches, OFM and DPM.

This value is tightly distributed around 4 in the results from the OFM approach, with a probability of $0.89$, 0.86, and 0.67 that four groups are needed to model $Y_1$, $Y_2$, and $Y_3$ respectively. For $Y_1$ and $Y_2$ there was a probability of 0.11 and 0.10 that only 3 groups were needed.

For $Y_2$ a small probability of 0.04 of 5 groups was also observed. For $Y_3$ the OFM model resulted in four or five groups, with a probability of 0.33 for the later. The large majority of iterations corresponded to a model with four occupied components for all three datasets under the OFM model.

The DPM model identified a larger number of occupied groups across all three datasets.
In comparison to the same value in the OFM, the DPM also exhibited much greater variability, resulting in a wider distribution over models with a varying number of occupied groups. For DPM however, the number of occupied groups is not expected to reflect the true value, and further investigation is required to examine the structure estimated by the two approaches.

\subsection{Pairwise allocation similarities} \label{sec:Result_2}
\begin{figure}[htb!]
\centering
 \includegraphics[width=\textwidth]{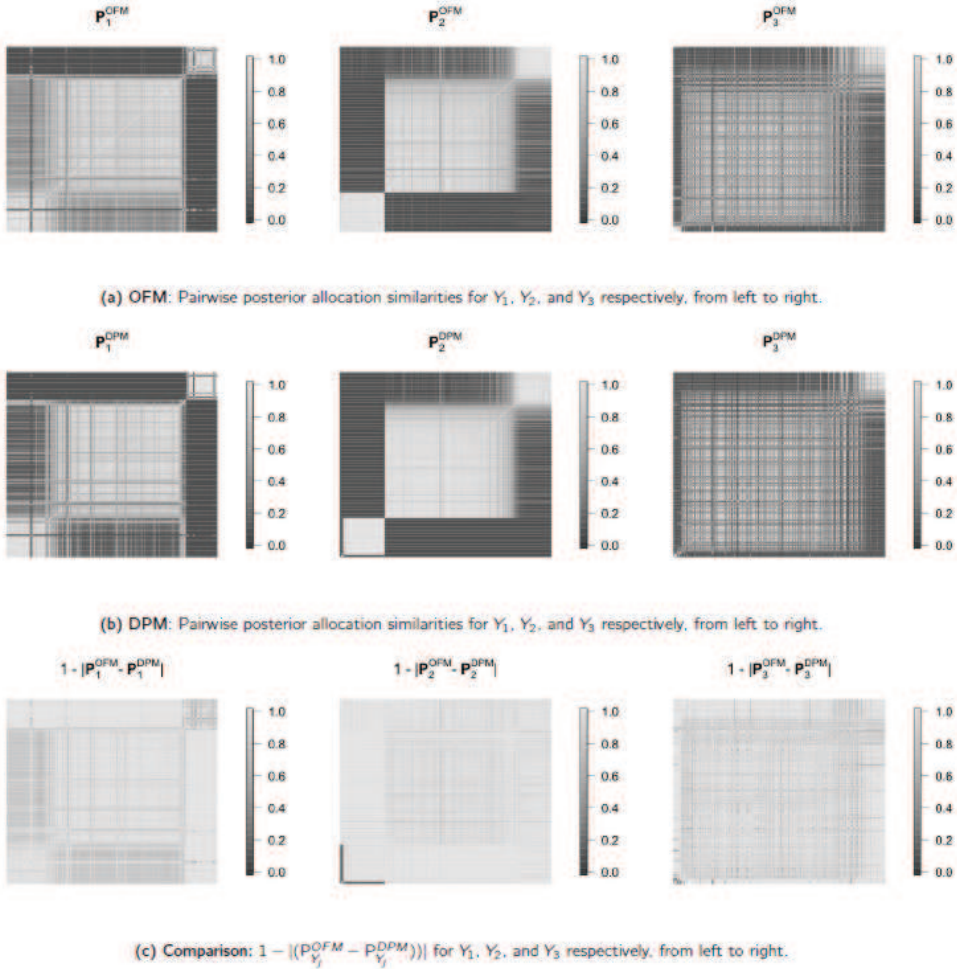}
\caption[Pairwise posterior allocation similarities.]{Plot of pairwise posterior allocation similarities for OFM and DPM ordered on both axes by the first dimension (PC 1) for each dataset ($Y_1$, $Y_2$, $Y_3$). a) OFM, b) DPM, c) Comparison.}
 \label{Figure_3}
\end{figure}

 Objective 2 is to explore whether the posterior membership of observations to components is similar under the OFM and DPM models. This is  explored in terms of the similarity of the pairwise posterior allocations. For each dataset $j=1,2,3$, a $n\times n$ matrix containing the proportion of MCMC iterations in which each pair of observations is allocated to the same group is computed. The estimated pairwise posterior allocation probability matrices thus created are denoted $\mathbf{P}_{Y_j}^{OFM}$ and $\mathbf{P}_{Y_j}^{DPM}$ to indicate the relevant method and dataset. The results are depicted in Figure \ref{Figure_3} (a) and (b), ordered by the first dimension of each $Y_j$; light areas of the plots correspond to large probabilities. In Figure \ref{Figure_3}(c), $1-|(\mathbf{P}_{Y_j}^{OFM} - \mathbf{P}_{Y_j}^{DPM})|$ is included to illustrate their differences more clearly; here large differences between the matrices correspond to dark areas.

The results indicate that both approaches sample a very similar posterior space, in that the estimated relationships between observations are quite consistent between the OFM and DPM models. In the case of $Y_1$ and $Y_2$ this was almost indistinguishable for most observations, while for $Y_3$, some structural differences exist, reflecting larger uncertainty in the clustering of these data. Direct interpretation is limited as the data are multivariate and corresponds to principal components.

For $Y_1$ and $Y_2$, three clusters are delineated by both approaches, visible in this plot as blocks of observations with a large probability of being allocated together. In both datasets, a large cluster near the median of the observations (dimension 1) is surrounded by two smaller groups.  The results of $Y_3$ are less interpretable, as ordering on the first dimension does not appear to capture the structure of the clusters well for this dataset. When comparing these directly ($3^{rd}$ plot of Fig.\ref{Figure_3}(c), the differences between the two appears to be slightly larger than for $Y_1$ and $Y_2$, but this is quite even and does not indicate any significant structural differences.

\subsection{Optimal partitioning} \label{sec:Result_3}
Objective 3 is to compare the composition of the components if a single model representing the optimal partition of the data into clusters is chosen. For the OFM approach, the chosen optimal model is the configuration with the most frequently reported number occupied components (as reported in Table \ref{Table_1}), as was done by \citet{VanHavre2015}. For the DPM approach, this is the PEAR estimate of the clustering, as discussed in Section \ref{sec:NPmodel} \citep{fritsch2009improved}.

\begin{table}[!htb]
\centering
\begin{tabular}{|c|cc|cc|cc|}
\hline
\textbf{Number of} & \multicolumn{2}{c}{$\mathbf{Y_1}$}&\multicolumn{2}{c}{$\mathbf{Y_2}$}&\multicolumn{2}{c|}{$\mathbf{Y_3}$}\\
\cline{2-7}
\textbf{occupied clusters} & \textbf{OFM} & \textbf{DPM} & \textbf{OFM} & \textbf{DPM} & \textbf{OFM} & \textbf{DPM}\\
\hline
1&--&--&--&--&--&--\\
2&--&--&--&--&--&--\\
3&0.106&--&0.100&--&--&--\\
4&0.894&0.270&0.862&--&0.672&--\\
5&--&0.458&0.04&0.867&0.328&--\\
6&--&0.201&--&0.131&--&--\\
7&--&0.065&--&0.002&--&0.073\\
8&--&0.006&--&--&--&0.679\\
9&--&--&--&--&--&0.238\\
10&--&--&--&--&--&0.010\\
\hline
\end{tabular}
\caption[Posterior distribution of the number of occupied components by model and dataset]{Posterior distribution of the number of occupied components by model and dataset, expressed as a proportion of the number of MCMC iterations.}
\label{Table_1}
\end{table}

The composition of each cluster, for each model,  is summarised in Table \ref{Table_2}. For the OFM, the majority of the observations were clustered in three large groups, and one small fourth cluster. The results of the DPM are similar for $Y_1$ and $Y_2$; the first three clusters are very close in size, the main difference being the presence of an extra group with a single observation. The structure of the inferred clusters appears to be quite different for $Y_3$ under the DPM, the model is composed of one large group of 195 observations, with a long tail of seven more clusters of decreasing size.


\begin{table}[htb!]
\centering
\begin{tabular}{|ccc|ccc|ccc|}
\hline
\multicolumn{9}{|c|}{\textbf{OFM}}\\
 \multicolumn{3}{|c}{$\mathbf{Y_1}$} &  \multicolumn{3}{c}{$\mathbf{Y_2}$} & \multicolumn{3}{c|}{$\mathbf{Y_3}$} \\\hline
 \textbf{k}  & \textbf{Count} & \textbf{\%} &\textbf{k} & \textbf{Count} & \textbf{\%} &  \textbf{k}   & \textbf{Count} &   \textbf{\%}  \\ \hline
1   &  101  & 52.60 \% &1 &	122  & 57.82\%&   1   &     176  & 50.57 \% \\
2   &  59   & 30.73 \% &2 &	44   &  20.85 \%&   2   &  72     &  20.69  \% \\
3   &  25   & 13.02 \% &3 &	42   & 19.91\%&   3   &  79     & 22.70  \% \\
4   &  7   &  3.65 \% &4 &   3   & 1.42\% &           4   & 21      & 6.03   \% \\
\hline
\multicolumn{9}{|c|}{\textbf{DPM}}\\
 \multicolumn{3}{|c}{$\mathbf{Y_1}$} &  \multicolumn{3}{c}{$\mathbf{Y_2}$} & \multicolumn{3}{c|}{$\mathbf{Y_3}$} \\\hline
 \textbf{k}  & \textbf{Count} & \textbf{\%} &\textbf{k} & \textbf{Count} &\textbf{ \%} &  \textbf{k}   & \textbf{Count} &  \textbf{ \%  }\\ \hline
1   & 112   & 58.33 &1&  125& 59.24&1& 195& 56.03\\
2   & 47    & 24.48 &2& 41& 19.43&2 &54& 15.52\\
3   & 25    & 13.02 &3& 42& 19.91&3 &42& 12.07\\
4   & 7     & 3.65 &4& 2& 2&4 &31& 8.91\\
5   & 1     & 0.52 &5& 1& 1& 5&13& 3.74\\
6   & --    & --   &6& --& --&6 &10& 2.87\\
7   & --    & --   &7& --& --&7 &2& 0.57\\
8   & --    & --   &8& --& --&8 &1& 0.29\\
\hline
\end{tabular}
\caption[Frequencies of cluster membership for a single model chosen under OFM and PDM]{Frequencies of cluster membership, as determined by the optimal partition under each model, by dataset.  The occupied clusters for each partition are given in decreasing order by frequency}
\label{Table_2}
\end{table}

\begin{figure}
  \centering
   \includegraphics[width=\textwidth]{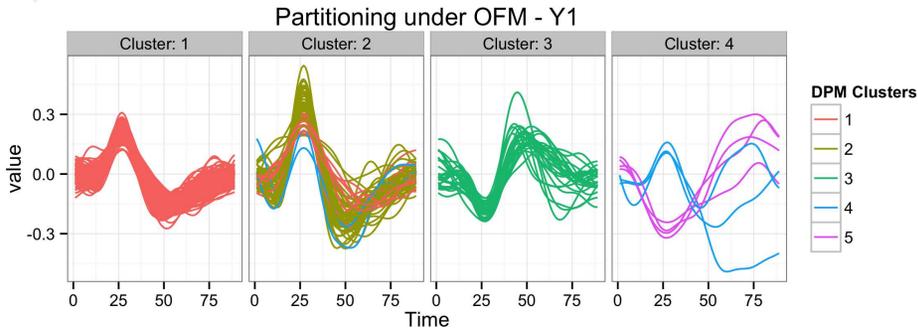}
  \caption[Inferred optimal partitions for $Y_1$ under the OFM]{Inferred optimal partitions for $Y_1$ under the OFM (one cluster per frame).  The clusters inferred under the DPM are represented by different colours, to indicate their composition/relationship with respect to clusters inferred by the OFM. }
  \label{Figure_4}
  \end{figure}

\begin{figure}
  \centering
   \includegraphics[width=\textwidth]{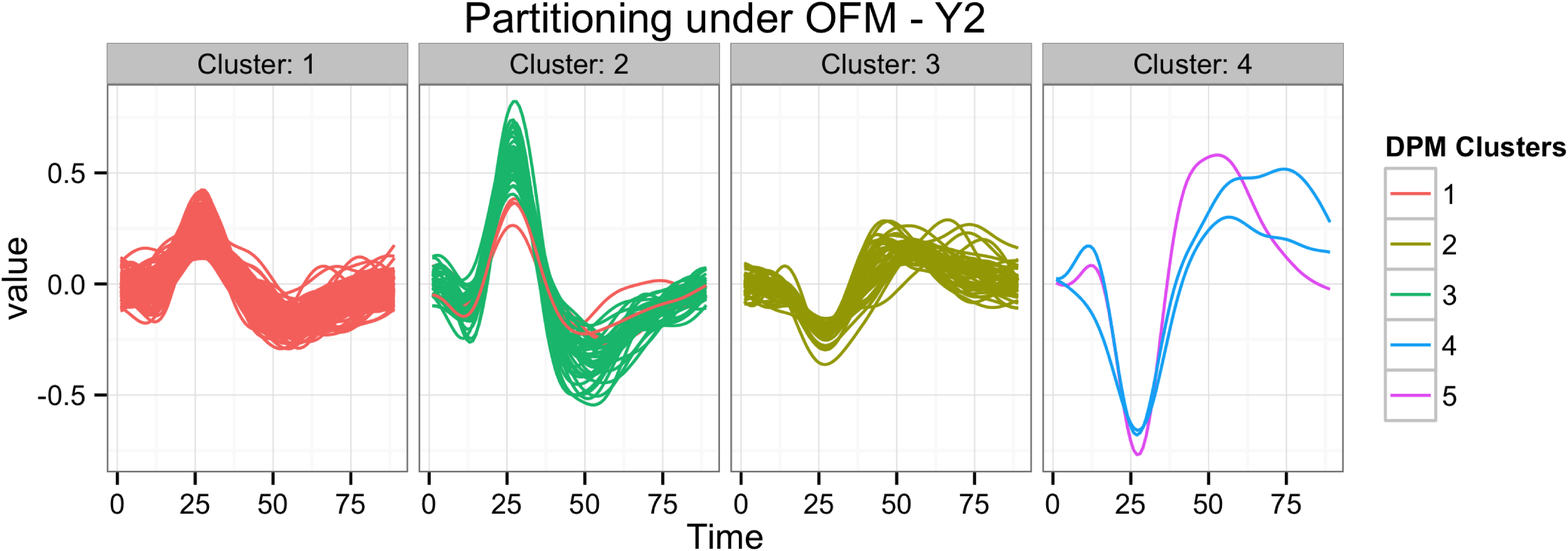}
  \caption[Inferred optimal partitions for $Y_2$ under the OFM]{Inferred optimal partitions for $Y_2$ under the OFM (one cluster per frame).  The clusters inferred under the DPM are represented by different colours, to indicate their composition/relationship with respect to clusters inferred by the OFM. }
  \label{Figure_5}
  \end{figure}

\begin{figure}
  \centering
  \includegraphics[width=\textwidth]{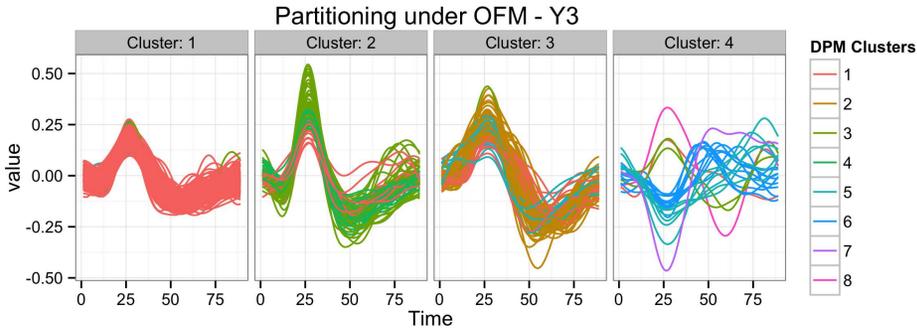}
  \caption[Inferred optimal partitions for $Y_3$ under the OFM]{Inferred optimal partitions for $Y_3$ under the OFM (one cluster per frame).  The clusters inferred under the DPM are represented by different colours, to indicate their composition/relationship with respect to clusters inferred by the OFM. }
  \label{Figure_6}
  \end{figure}

Figures \ref{Figure_4}, \ref{Figure_5} and \ref{Figure_6}  allow for the direct comparison of the inferred clusters. The spikes are shown plotted according to the groups found by the OFM, with each spike coloured according to their optimal clustering under the DPM. The plots are ordered according to the number of observations allocated in the OFM clusters, so that group 1 is the largest OFM group and 4 the smallest, as in Table \ref{Table_2}.  The inverse plots of the spikes sorted by the DPM clusters and coloured according to the OFM clusters can be found in the Supplementary Material (see \nameref{Figure_1_SuppInfo},  \nameref{Figure_2_SuppInfo}, and  \nameref{Figure_3_SuppInfo}).

For all three datasets, the three largest OFM clusters correspond closely with different PDM clusters; all spikes in clusters 1  and 3 of $Y_1$ and $Y_2$ (see Figures \ref{Figure_4} and \ref{Figure_5}) are from a single DPM group. In a few of the large groups, a small number of spikes are allocated originating from other DPM clusters, as can be seen by the small number of red and blue lines in cluster 2  for $Y_1$ (Figure \ref{Figure_4}).

The results of $Y_3$ in Figure \ref{Figure_6} indicate larger differences between the clusterings found by OFM and DPM. The OFM cluster 1 corresponds closely to DPM cluster 1, indicated by red curves. This is also true for third largest OFM group, which contains mostly spikes from DPM group 2 (with a few extra observations from the red DPM group). OFM group 2 is made of two different DPM clusters, those indicated by the dark blue and pink lines.

The composition of the smallest clusters (OFM cluster 4) is different from the other three groups, and appears to contain spikes of widely varying shapes which do not fit well into the other groups. For $Y_1$, this contains 7 observations, all from the two smallest DPM clusters. In $Y_2$, the smallest OFM cluster contains only 3 observations, again from the two smallest DPM components.  In $Y_3$ the fourth OFM cluster is larger and contains 21 observations, but is again made up mostly of spikes allocated to the small, variable clusters found by the DPM.


\newpage

\section{Discussion} \label{sec:discussion}
Performing a comparison of the two approaches for mixtures with an unknown number of components, one parametric and one non-parametric, provides insights on the impact of the underlying assumptions involved in both methods. The two methods, OFM and DPM are used to fit similar multivariate Gaussian finite mixture models. The small differences in these reveal how uncertainty is handled under the two approaches, both in the posterior distribution of occupied cluster and in the composition of the clusters identified. This understanding is crucial for the future applications of mixture models that seek to describe complex data. 

The choice of method will depend on the goal of the analysis. Both methods are effective at identifying high probability clusters made up of spikes with similar trajectories, effectively capturing the bulk of the variability in the spike shapes. The uncertainty in the clustering in both methods is caused by the presence of a small number of spikes in each dataset which tend to not be allocated in these larger clusters. In the DPM, these are captured with a number of small clusters with small posterior weights, reflecting a range of potential models explaining the components for which few observations are present. Thus the number of occupied groups in the MCMC tends to vary and be greater than may be supported by the data, even though most observations are allocated to a small number of occupied components. In this manner, the DPM approach is able to capture fine details in structure pertaining to potentially very small clusters, and reflects their lack of strong support with greater variability in these small clusters.

The OFM treats the uncertainty in the clustering differently, as the prior strongly discourages the posterior from placing mass on clusters which are not strongly supported by observations. The observations which do not fit into the high probability groups (with large weights), are combined into a single group with a large covariance, capturing the outliers with the addition of a multivariate Gaussian noise component. This prevents interpretation of the smallest clusters, as may be possible with the results of the DPM if the small components can be justified.

Well defined clusters are, on the whole, easy to identify using the methods considered and are expected to be so using any number of other approaches for unsupervised clusters.
The differences in the results occurred where data was insufficient to overcome the suggestion of the priors. A clear understanding of exactly how such uncertainty is handled in both the OFM and DPM is crucial to their effective use in practice, allowing the future analyst to choose an approach suited to the situation and goal of the research problem at hand and interpret the results. 

When the aim is to obtain a sparse clustering strongly supported by the data, as well as to identify Gaussian noise, the OFM approach is ideal and requires few inputs or decisions on the part of the analyst to obtain the result. In cases where a `hard' clustering is too restrictive to explain complex data however, the DPM is able to explore a finer, more intricate underlying structure, and as such can estimate a larger number of small components which are less likely, but possible given the data. In a sense, the OFM provides information on the \emph{probable} set of models which explain the data, while the DPM provides informations about all \emph{possible} models, producing a richer but slightly more cumbersome result.

\newpage
\section{Supporting Information} 
Supporting Information: Additional information for this article is available.
\subsection*{Supplementary Figure 1}\label{Figure_1_SuppInfo}
   \includegraphics[width=\textwidth]{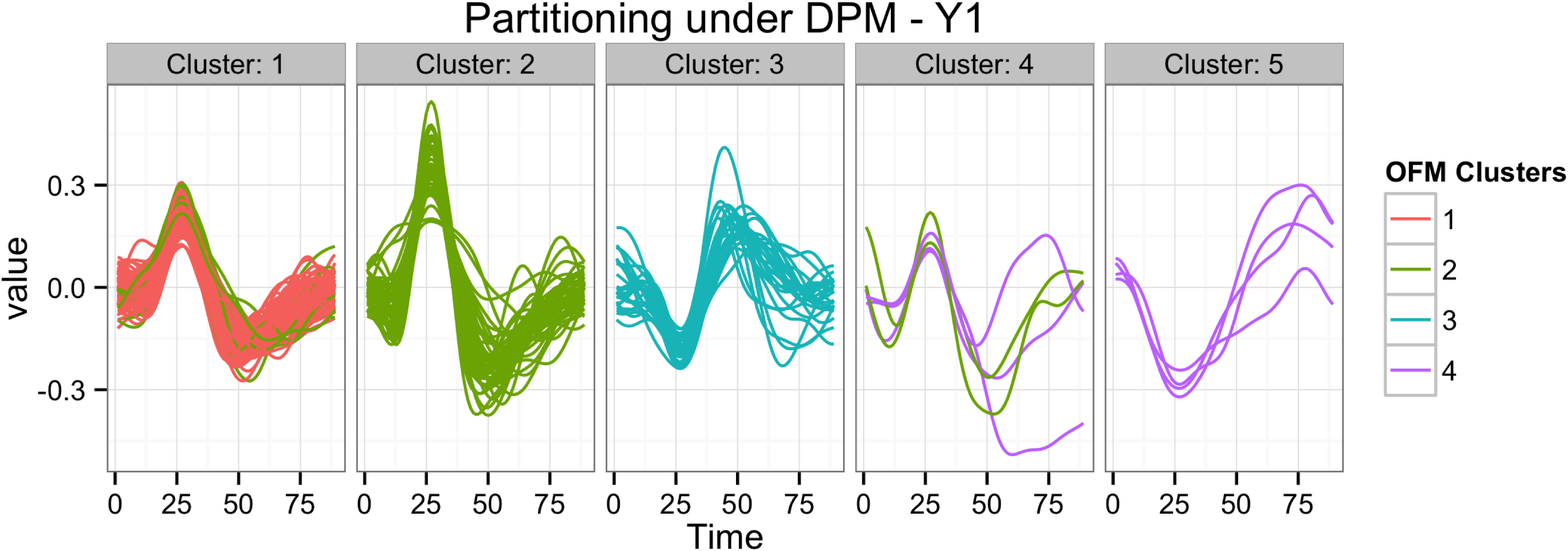}
{\bf Inferred optimal partitions for $Y_1$ under the DPM}  
The clusters inferred under the OFM are represented by different colours, to indicate their composition/relationship with respect to clusters inferred by the DPM.

\subsection*{Supplementary Figure 2}\label{Figure_2_SuppInfo}
   \includegraphics[width=\textwidth]{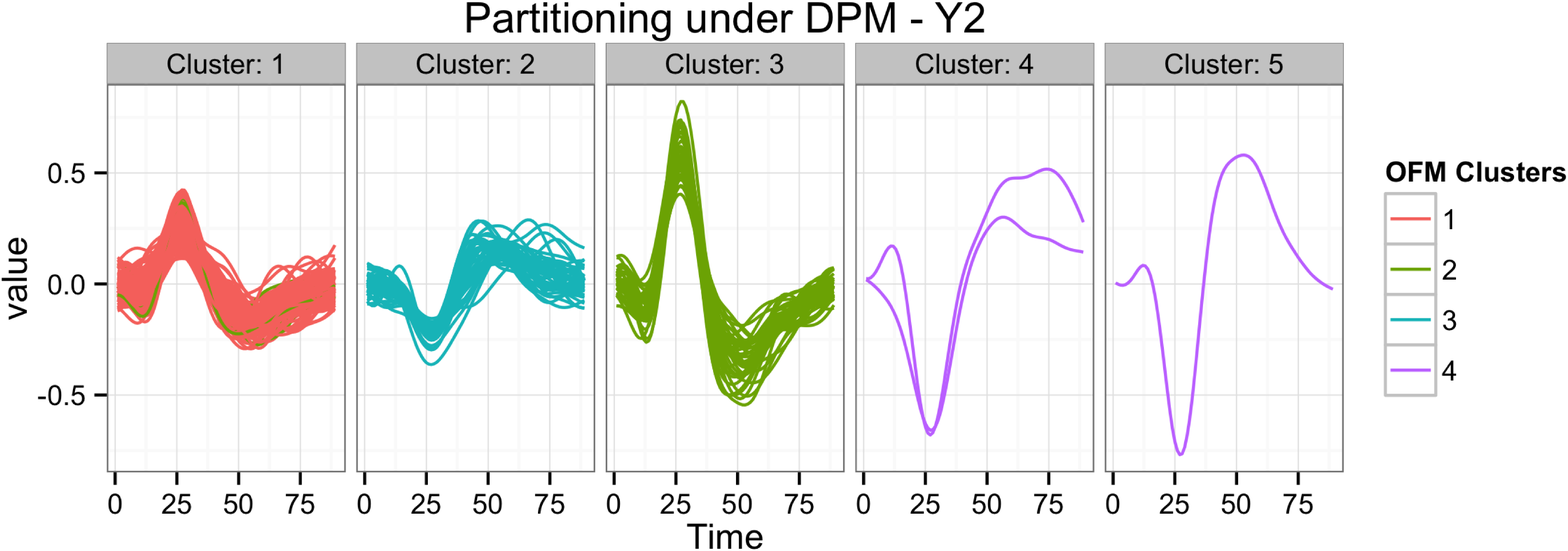}
{\bf Inferred optimal partitions for $Y_2$ under the DPM} The clusters inferred under the OFM are represented by different colours, to indicate their composition/relationship with respect to clusters inferred by the DPM.

\subsection*{Supplementary Figure 3}\label{Figure_3_SuppInfo}
   \includegraphics[width=\textwidth]{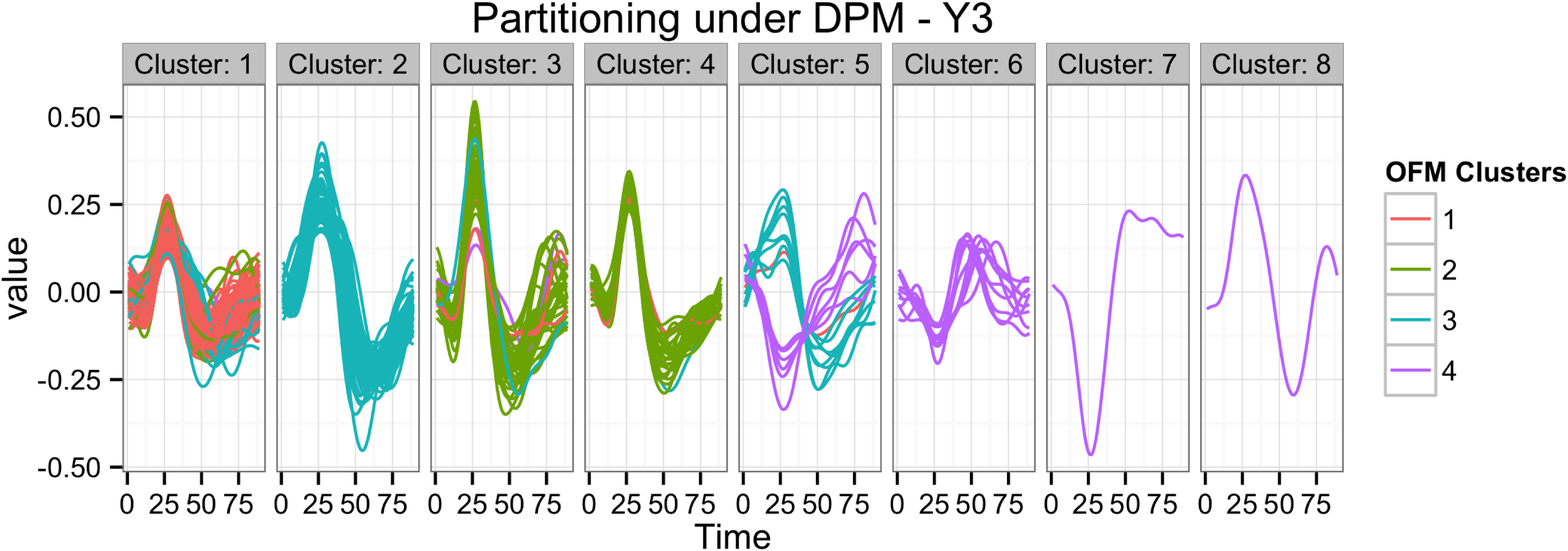}
{\bf Inferred optimal partitions for $Y_3$ under the DPM}
The clusters inferred under the OFM are represented by different colours, to indicate their composition/relationship with respect to clusters inferred by the DPM.

\subsection*{Supplementary File 1}\label{Code_1_SuppInfo}
{\bf File for implementation of Overfitted Finite Mixture model (OFM).} The file contains R code which allows for the estimation of a Multivariate Gaussian OFM using Markov chain Monte Carlo (MCMC).

\subsection*{Supplementary File 2}\label{Code_2_SuppInfo}
{\bf Files for implementation of Dirichlet Process mixture model (DPM).} The files contained in this folder allow for the estimation of a Multivariate Gaussian DPM using Markov chain Monte Carlo (MCMC).

\paragraph{Download:} Please download Supplementary File 1 and 2 by
\href{https://drive.google.com/file/d/0B3FRvR7x47lIYkFobTBYNXBteDQ/view?usp=sharing}{clicking here}

\end{document}